\documentclass[11pt]{article}

\usepackage[T1]{fontenc}
\usepackage[margin=1in]{geometry}
\usepackage{longtable}
\usepackage{booktabs}
\usepackage{array}
\usepackage{hyperref}
\usepackage{graphicx}

\hypersetup{
  colorlinks=true,
  linkcolor=black,
  citecolor=black,
  urlcolor=blue
}

\title{Workplace Demands and Emotional Expression Among Early Childhood Educators: A Computational Analysis of Professional Online Discourse}

\author{
Hailong Jiang\\
Department of Computer Science, Information, \& Engineering Technology\\
Youngstown State University\\
\texttt{hjiang@ysu.edu}
}

\date{April 2026}

\begin{document}

\maketitle
\begin{abstract}
Early childhood educators work in settings marked by heavy regulation, emotional labor, staffing instability, and low pay. Although these conditions are well documented in survey research, less is known about how they appear in the day-to-day language educators use in peer spaces. This study examined 7,506 posts from \texttt{r/ECEProfessionals}, a large online community used by early childhood education practitioners. Using a structured computer-assisted thematic coding workflow and transformer-based emotion classification, posts were organized into 15 themes and then mapped onto an adapted Job Demands--Resources (JD-R) framework. Across the corpus, 56.7\% of posts centered on demands when task-level and core job demands were combined, compared with 33.6\% focused on resources and 9.6\% on career conditions. Emotion estimates showed a broadly neutral tone overall, but fear emerged as the most prominent non-neutral emotion. Demand-related categories also showed more sadness and anger than resource-related categories. These findings suggest that professional online discourse in early childhood education reflects a work environment structured more around strain than support. The study offers a practical framework for examining how occupational conditions are discussed and emotionally experienced in large-scale professional text.
\end{abstract}

job demands--resources; emotion analysis

\section{Introduction}

Workplace well-being in education has usually been studied through surveys, interviews, and institutional reports. These approaches have been essential for documenting burnout, stress, and turnover, but they capture experience retrospectively and within predefined measurement frames (Agyapong et al., 2022; Grant et al., 2019). Professional online communities offer a useful complement. In these spaces, educators describe problems, seek advice, and interpret workplace events in their own language, often close to the time those events occur (De Choudhury et al., 2013). As a result, online discourse can reveal how occupational strain is collectively framed, not only how it is measured after the fact.

Early childhood education (ECE) is a particularly important setting for this kind of inquiry. ECE work combines teaching, caregiving, emotional support, behavior management, documentation, and family communication. At the same time, the sector is marked by staff shortages, low pay, high turnover, and strong regulatory demands (Wiltshire, 2024; Stein et al., 2022). Recent studies continue to describe elevated stress, depressive symptoms, and emotional exhaustion among early childhood educators, as well as clear links between working conditions and intentions to leave the field (Fuentes-Vilugron et al., 2025; Grant et al., 2019; Madigan \& Kim, 2021). Emotional labor is also central to the profession. Educators are expected to remain calm, responsive, and professional even in emotionally taxing situations, and these demands may accumulate when organizational support is weak (Purper et al., 2023; Carey \& Sutton, 2024; Dickerson et al., 2025).

These concerns matter for more than workforce retention alone. Teacher strain can shape the emotional climate of classrooms and the quality of daily interactions available to young children. Prior work has linked educator stress and burnout to lower-quality interactions and to physiological stress processes in educational settings (Jogi et al., 2023; Schlueter et al., 2024). For that reason, understanding how educators experience and talk about work is not only a labor issue, but also an educational one. If chronic strain becomes normalized in the profession, it may affect both adult well-being and the environments in which children learn and develop.

The Job Demands--Resources model provides a useful framework for organizing these conditions. The model distinguishes between job demands, which require sustained physical or psychological effort, and job resources, which help employees meet work goals, reduce strain, or support growth (Demerouti et al., 2001; Bakker \& Demerouti, 2007). In educational settings, demands may include workload, role conflict, health exposure, and difficult relationships with families, whereas resources may include supportive leadership, collegial support, training, autonomy, and effective communication systems (Tummers \& Bakker, 2021). In ECE specifically, poor working conditions have been linked to burnout and turnover, while stronger organizational and relational supports appear to buffer risk (Grant et al., 2019; Carey \& Sutton, 2024; Oosterhoff et al., 2020).

Although the ECE stress literature is well established, most studies rely on surveys or interview data. Less is known about how educators themselves narrate workplace strain in informal, peer-oriented settings. This gap matters because the discourse of professional communities may reveal not only what problems are present, but also which problems feel most immediate, discussable, or emotionally charged. Reddit and related platforms have increasingly been used for research on stress, affect, mental health, and public perceptions of health-related topics because they provide large volumes of naturally occurring, relatively candid text (Dan et al., 2025; Zhu et al., 2023; Zhu et al., 2025a). Advances in natural language processing now make it possible to study large corpora without reducing analysis to simple word counts. Structured thematic workflows can support consistent classification across thousands of posts, and transformer-based emotion models can estimate patterns in affective language at scale (Ahmed et al., 2025; Hartmann, 2022; Zhu et al., 2025b).

Professional online communities are especially useful in occupations where workers may hesitate to speak candidly in employer-controlled settings. In pseudonymous forums, educators can ask for advice, describe conflicts, and test interpretations of policy or workplace behavior with less reputational risk than they would face in their own centers. Such discourse is not a neutral or representative sample of the workforce, but it can still illuminate what concerns feel urgent enough to bring to peers. For ECE research, this matters because the field is often studied through formal instruments, while the everyday language of educators remains less visible in the literature.

The present study uses these tools to examine discussion within \texttt{r/ECEProfessionals}, a large online community oriented toward ECE practice. Rather than treating online discourse as a substitute for validated measures of well-being, the study treats it as a complementary record of how work is talked about in situ. Two goals guided the analysis. The first was to identify the major topics that organize professional discussion in this community. The second was to examine whether affective expression differs across discourse centered on demands, resources, and longer-term career conditions. To do so, posts were first assigned to a set of primary themes developed through a structured computer-assisted coding workflow. These themes were then organized within an adapted JD-R framework, and emotion probabilities were aggregated across categories.

This focus is especially relevant in ECE because many central pressures are organizational and relational rather than purely instructional. Posts about staffing, management, parent expectations, and compensation can reveal how educators interpret the structure of their work, where they locate support, and what kinds of uncertainty remain unresolved. Examining discourse at scale therefore provides a way to study not only individual distress, but also the everyday social meaning of professional demands.

The study addresses three questions: What themes characterize professional discourse in this ECE community? What is the overall emotional profile of that discourse? How do emotion patterns vary across job demands, job resources, and career-related conditions? By focusing on discourse rather than survey response, the article extends ECE well-being research into a digital setting where practitioners publicly negotiate everyday work realities. It also offers a practical framework for using large-scale text data to study the occupational climate of care and education work.

\section{Method}

\subsection{Data Source and Corpus Construction}

The dataset consisted of 7,506 posts collected from \texttt{r/ECEProfessionals}, a public Reddit community used by early childhood education practitioners. Posts were retrieved from publicly available archives and filtered to retain original posts with substantive text. Comments were excluded so that each observation represented a self-contained discussion prompt rather than an exchange thread. Available metadata included the post title, body text, timestamp, and engagement indicators. Author identifiers were removed before analysis.

Each post was treated as a single analytic unit. Titles and body text were combined to preserve context, since many posts used the title to signal the problem and the body to elaborate it. The resulting corpus was designed to capture how educators framed work-related concerns, requests for advice, and reflections on practice in a naturally occurring peer environment.

The study analyzed discourse at the level of posts rather than users. This choice kept the focus on topics and emotional tone within professional communication rather than on individual posting histories. It also aligned with the substantive aim of the study, which was to describe the structure of discussion in the community rather than to profile specific participants or identify high-frequency users.

Because the corpus was drawn from a public online forum, analytic decisions also weighed privacy and traceability concerns. The study did not attempt to reproduce user histories or identify individual posters. Direct quotations were avoided in the present manuscript to reduce the likelihood that search engines could reconnect excerpts to source accounts. This approach was intended to balance the public nature of the data with the sensitivity of work-related and emotional disclosures.

\subsection{Thematic Coding Workflow}

To make corpus-wide classification manageable while maintaining analytic transparency, the study used a staged computer-assisted coding workflow. First, a development sample of 50 posts was reviewed to identify recurrent concerns and linguistic patterns. Longer posts were segmented at the paragraph level for memoing and preliminary open coding. A large language model was used at this stage as an analytic support tool to summarize segments and propose candidate codes, consistent with recent work using large language models to analyze emotional and contextual patterns in social media discourse (Zhu et al., 2025c). These outputs were not treated as final interpretations. Instead, they were compared across cases and consolidated into a codebook of 15 themes, each defined in operational terms.

The codebook was then used to assign one dominant theme to each post in the full corpus. This procedure separated exploratory code generation from final thematic application and reduced the risk that a single automated pass would determine the analytic structure. Theme definitions were refined through iterative comparison between model-assisted summaries and close reading of posts so that labels remained grounded in the corpus rather than in generated output alone. The goal of the workflow was descriptive classification rather than discovery of latent causal mechanisms. For that reason, the thematic results are interpreted as structured summaries of discourse content in this community.

Assigning a single dominant theme necessarily simplified some posts that involved multiple concerns. In practice, however, many posts still had a clear center of gravity, such as a conflict with management, a licensing concern, a compensation question, or a classroom-management problem. A single-theme approach was retained because it supported a consistent corpus-wide comparison with the JD-R framework. It also prevented frequent double counting of broad complaints that touched several related issues at once.

\subsection{JD-R Organization}

After primary themes were identified, they were organized within an adapted JD-R framework. Four structural categories were used. Task demands referred to routine instructional and classroom-management work. Core job demands referred to broader strains such as regulation, conflict, burnout, illness exposure, emotional labor, and material scarcity. Job resources referred to supports such as leadership, staffing, communication systems, inclusion practices, and advocacy channels. Structural career conditions referred to compensation, employability, and longer-term career stability.

This four-part organization was used to interpret whether emotional expression differed across forms of occupational strain and support. It also made it possible to distinguish between day-to-day pedagogical work and broader structural pressures that extend beyond the classroom.

\subsection{Emotion Classification}

Emotional expression was estimated using the \texttt{j-hartmann/emotion-english-}\\\texttt{distilroberta-base} model (Hartmann, 2022). For each post, the model returned probabilities for seven categories: sadness, disgust, anger, neutral, fear, surprise, and joy. Probability scores were retained rather than converted to hard labels so that emotional profiles could be averaged across groups without forcing each post into a single emotion. Because the model was trained on general English rather than ECE-specific discourse, results were interpreted as comparative indicators of affective language rather than as precise measures of speaker emotion.

Emotion probabilities were aggregated in two ways. First, scores were averaged across the entire corpus to describe the overall emotional landscape. Second, scores were averaged within each JD-R category to compare affective patterns across task demands, core job demands, job resources, and structural career conditions. Because the study was descriptive, the emphasis was on comparative patterning rather than causal inference.

No attempt was made to infer clinical states from the text. The emotion model was used only to estimate relative patterns in affective language at the group level. This distinction is important because online posts can contain frustration, humor, or practical concern without constituting evidence of mental disorder or impairment.

\section{Results}

\subsection{Thematic Distribution}

The 7,506 posts were distributed across 15 primary themes. The most common theme was classroom practice, pedagogy, and behavior management (16.40\%), followed closely by leadership, management, and workplace culture (15.38\%). Other frequent themes included safety, regulatory compliance, and legal risk (9.63\%), compensation and career pathways (9.60\%), and workforce stress, burnout, and turnover (9.36\%). Lower-frequency themes, such as inclusion and cultural competence (1.69\%) and advocacy and systemic supports (1.57\%), appeared less often but still represented recognizable parts of the discourse.

When the themes were grouped by the adapted JD-R framework, demand-oriented discourse dominated the corpus. Task demands accounted for 16.40\% of posts, and core job demands accounted for 40.29\%. Combined, 56.69\% of the corpus focused on demands. Job resources accounted for 33.64\% of posts, while structural career conditions accounted for 9.60\%. In other words, educators discussed strain more often than support, and long-term career issues occupied a visible but smaller share of the conversation.

This pattern suggests that the online community functioned less as a space for celebratory professional exchange and more as a venue for processing everyday pressures. At the same time, the strong presence of leadership and workplace culture among resource-oriented posts indicates that institutional support remained a central concern rather than a peripheral one.

The thematic pattern also highlights an important asymmetry in the discourse. Resource-oriented posts were common, but they were distributed across several different topics, including management, staffing, communication, inclusion, and advocacy. By contrast, demand-related content was more concentrated around structurally difficult issues such as compliance, burnout, parent conflict, and compensation. This suggests that support in the field may be experienced as fragmented, whereas strain is experienced as cumulative and interconnected.

The distribution of the top themes reinforces this interpretation. The five most frequent themes alone accounted for just over 60\% of the corpus, and four of those five were directly tied to structural working conditions rather than to instructional content alone. Even where posts focused on teaching practice, they were often embedded in wider discussions about staffing, behavior expectations, or institutional constraints. Taken together, the thematic pattern suggests that educators do not experience classroom work in isolation from employment conditions and organizational systems.

\begin{longtable}{p{0.18\textwidth}p{0.50\textwidth}r r}
\caption{Distribution of Themes Within the Adapted JD-R Framework} \\
\toprule
JD-R Category & Theme & $n$ & \% \\
\midrule
\endfirsthead
\toprule
JD-R Category & Theme & $n$ & \% \\
\midrule
\endhead
Task demand & Classroom practice, pedagogy, and behavior management & 1,232 & 16.40 \\
Resource & Leadership, management, and workplace culture & 1,155 & 15.38 \\
Core job demand & Safety, regulatory compliance, and legal risk & 723 & 9.63 \\
Structural career condition & Compensation, career pathways, and employability & 721 & 9.60 \\
Core job demand & Workforce stress, burnout, and turnover & 703 & 9.36 \\
Core job demand & Parent-staff boundaries, conflict, and expectations & 633 & 8.43 \\
Resource & Staffing, training, and role readiness & 509 & 6.78 \\
Core job demand & Health, illness exposure, and public-health pressures & 356 & 4.74 \\
Core job demand & Emotional labor, professionalism, and respect & 349 & 4.65 \\
Resource & Communication, documentation, and parent engagement practices & 336 & 4.47 \\
Resource & Access to supports, special needs, and early intervention & 282 & 3.75 \\
Core job demand & Resource constraints, cost-quality tensions, and teacher-paid expenses & 167 & 2.22 \\
Resource & Inclusion, cultural competence, and identity respect & 127 & 1.69 \\
Resource & Advocacy, escalation channels, and systemic supports & 118 & 1.57 \\
Core job demand & Parent expectations, safety trade-offs, and risk tolerances & 95 & 1.26 \\
\bottomrule
\end{longtable}

\subsection{Overall Emotional Landscape}

Across the full corpus, neutral language was the most prevalent emotional category, with a mean probability of approximately .31. This indicates that many posts were framed in informational or experience-sharing terms rather than in highly expressive language. Fear was the most prominent non-neutral emotion, with a mean probability of approximately .18. Sadness, anger, and disgust appeared at moderate levels, whereas joy remained comparatively low.

This profile suggests that the community was not dominated by overtly dramatic or celebratory discourse. Instead, the emotional tone was one of persistent concern embedded within practical discussion. The prominence of fear is noteworthy because it points to uncertainty and vulnerability rather than simple irritation alone.

\begin{figure}[t]
\centering
\fbox{%
    \includegraphics[width=0.88\linewidth]{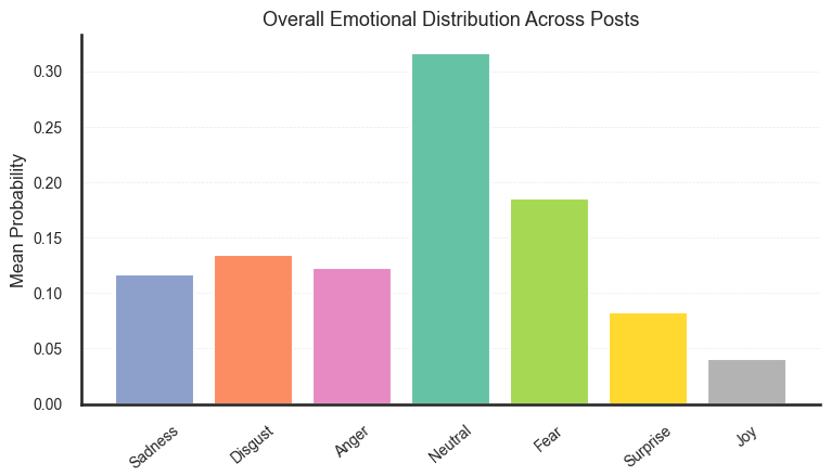}%
}
\caption{Overall emotion distribution across the corpus.}
\label{fig:emotion_distribution}
\end{figure}

\subsection{Affective Patterns Across JD-R Categories}

Emotion patterns varied across the four JD-R categories. Core job demand posts showed the strongest concentration of fear and also higher sadness and anger than resource-oriented posts. Structural career condition posts, especially those focused on compensation and employability, also showed relatively high fear, consistent with economic uncertainty and career instability. By contrast, job resource posts were more neutral overall and showed lower negative affect. Task-demand posts, particularly those focused on everyday classroom practice, showed the highest neutrality of the four categories.

These differences suggest that not all frequently discussed topics are equally affectively charged. Routine pedagogical concerns were common, but they were often handled in a pragmatic register. More structural topics, such as burnout, regulation, illness exposure, and compensation, were discussed with greater emotional strain.

\begin{figure}[t]
\centering
\fbox{\parbox{0.88\linewidth}{\includegraphics[width=\linewidth]{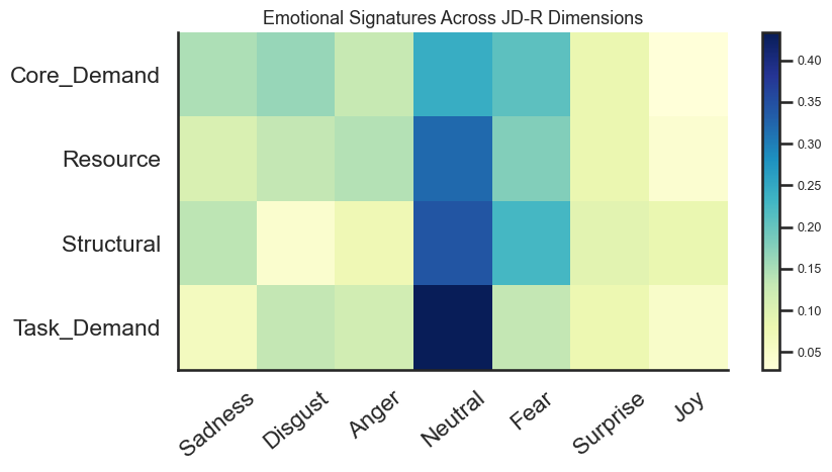}}}
\caption{Mean emotion probabilities across JD-R structural categories.}
\end{figure}

\section{Discussion}

This study examined how early childhood educators discussed work in a large online professional community and how emotional expression varied across different kinds of occupational content. The clearest finding was structural imbalance. More than half of the posts centered on demands when task demands and core job demands were combined, while roughly one-third focused on resources and fewer than one in ten focused on career conditions. This distribution is consistent with previous research showing that ECE work is shaped by chronic strain, limited support, and instability in the labor force (Wiltshire, 2024; Stein et al., 2022; Grant et al., 2019).

The emotional profile of the corpus deepens that interpretation. Although much of the discourse was neutral in surface tone, fear stood out as the most prominent non-neutral emotion. This matters because fear suggests a particular kind of occupational climate. Anger often signals protest, whereas fear can reflect uncertainty about safety, compliance, job security, professional legitimacy, and financial survival. The pattern observed here therefore points not only to dissatisfaction, but also to vulnerability. In ECE settings, such vulnerability may arise from high accountability, low control, exposure to illness, parent conflict, and economic precarity, all of which are well represented in the existing literature (Oosterhoff et al., 2020; Carey \& Sutton, 2024; Fuentes-Vilugron et al., 2025).

The contrast between categories is also informative. Posts about classroom practice were frequent but comparatively neutral, suggesting that many day-to-day teaching concerns are discussed as practical problems to solve. Resource-oriented posts were likewise more neutral than demand-oriented posts, which may indicate that discussions of support, staffing, training, and advocacy are relatively solution focused. By comparison, posts grouped under core job demands and career conditions were more affectively negative. This pattern is consistent with JD-R theory, which predicts that chronic demands carry psychological cost when they are not adequately offset by resources (Bakker \& Demerouti, 2007; Schaufeli \& Taris, 2014).

One especially notable finding is the prominence of leadership, management, and workplace culture as the largest resource-related theme. In many ECE settings, direct supervisors translate policy into daily practice, shape staff communication, and influence whether educators feel respected or expendable. The fact that management appeared so frequently in the discourse suggests that organizational climate is not a background factor. It is one of the main ways educators encounter either support or strain. This observation fits JD-R scholarship emphasizing the role of leadership in mobilizing resources and mitigating demands (Tummers \& Bakker, 2021).

For practice, the findings point to several areas of concern for program leaders and policy makers. First, the prominence of burnout, regulation, parent conflict, and compensation suggests that workforce well-being cannot be addressed through individual coping strategies alone. Structural supports matter. Staffing stability, clearer communication systems, responsive leadership, and more realistic compliance expectations are likely to affect educator well-being directly. Second, the visibility of compensation and career-pathway concerns indicates that financial and professional insecurity remain central to how educators understand the field. Third, because online professional communities capture naturally occurring discussion, they may provide a useful supplementary source of information for organizations seeking to monitor emerging stressors in the workforce.

These findings also have implications for how support is designed. Interventions focused only on mindfulness or resilience may help some educators, but they do not address the recurrent topics that dominated this corpus. If educators are consistently talking about staffing gaps, inconsistent management, parent boundary issues, and economic precarity, then the most meaningful interventions are likely to be organizational and policy based. Efforts to reduce unnecessary paperwork, clarify family communication protocols, protect planning time, create predictable staffing structures, and improve compensation may have broader effects than isolated wellness initiatives. In that sense, the discourse itself points toward where institutional attention is most needed.

At the same time, the relatively neutral tone of resource-oriented posts suggests that supportive infrastructures can shift conversation away from emotional overload and toward problem solving. This does not mean that resource-rich settings are free of stress. Rather, it suggests that educators may communicate differently when they perceive a pathway to action. In practical terms, leaders may benefit from paying attention not only to whether staff raise concerns, but also to the emotional register in which those concerns are voiced. A workplace dominated by apprehension is different from one in which problems are raised with confidence that help exists.

From a policy standpoint, the findings underscore how closely educator well-being is tied to the structure of ECE work. Discussions of compensation, employability, legal risk, and illness exposure were not marginal in the corpus; they were woven into the main fabric of professional talk. This is important because it challenges narratives that frame burnout mainly as an individual capacity problem. A field characterized by low wages, unstable staffing, high accountability, and emotional overload will continue to generate distress even when individual educators are highly committed and skilled. Policies that improve compensation, staffing ratios, paid time, access to substitutes, and administrative consistency are therefore likely to matter not only for retention, but also for the emotional climate of the profession.

The study also contributes methodologically. It shows that large-scale discourse analysis can be linked to an occupational theory that is already familiar in ECE and teacher well-being research. The value of this approach is not that it replaces surveys, interviews, or observational work. Rather, it offers a way to examine what practitioners choose to discuss spontaneously and how those discussions are emotionally inflected.

This methodological contribution is particularly useful for topics that are hard to observe directly at scale. Surveys can tell researchers whether educators report burnout or low support, but they often say less about the specific situations that organize concern in everyday professional life. Text-based analysis can help fill that gap by showing how occupational issues are clustered, what language practitioners use, and which topics attract the most emotionally charged discussion. Used carefully, such approaches can complement more traditional ECE research designs.

The findings also speak back to JD-R theory in a modest but meaningful way. In this corpus, demands did not appear only as isolated stressors. They clustered across institutional, relational, and economic domains, which is precisely the kind of accumulation the JD-R perspective highlights. At the same time, resources appeared less as broad protective states and more as practical sites where support is negotiated, judged, or repaired. This suggests that applying JD-R to digital discourse may help researchers see how educators themselves describe the balance, or imbalance, between strain and support in their work lives.

\section{Limitations and Future Work}

Several limitations should be considered. First, the corpus came from a single Reddit community. The users of this forum are self-selected, and the discourse may not represent the views of the broader ECE workforce. Second, the analysis focused on original posts rather than comments, which means that it captured initial problem framing more than full conversational exchange. Third, the thematic workflow relied in part on a large language model to support memoing and preliminary coding. Although the workflow was structured and transparent, automated assistance does not eliminate interpretive bias. Fourth, the emotion model was trained on general English emotion data rather than specifically on educator discourse, so some domain-specific language may have been simplified or misclassified. Finally, the study was descriptive and cross-sectional. It identifies patterns in discourse but does not establish causal relationships between working conditions and well-being.

An additional limitation is that the study prioritized broad thematic coverage over close reading of illustrative excerpts. This decision supported privacy protection and large-scale comparison, but it also means that some nuance in voice, irony, and context is inevitably compressed in the aggregate results. Future work could combine corpus-level analysis with smaller-scale qualitative interpretation to preserve more of that texture.

Future studies could strengthen this line of work by incorporating human validation of theme assignments, examining temporal shifts in discourse, comparing multiple educator communities, and integrating text-based findings with survey or interview data. Such extensions would help clarify whether the demand-heavy emotional profile observed here generalizes across contexts and over time.

Longitudinal work would be especially valuable. Online discourse is well suited to tracing how educators respond to policy shifts, seasonal illness patterns, labor shortages, or public controversies over curriculum and regulation. A time-sensitive design could show whether fear rises during moments of institutional uncertainty, whether resource-oriented discussion expands when supports improve, and whether some topics recur predictably across the academic year. That kind of analysis would move the field closer to understanding not only what educators discuss, but when particular forms of strain become most visible.

Cross-platform comparison would also strengthen interpretation. Educator talk on Reddit may differ from discussion in closed professional groups, unions, or platform-specific communities with different norms of disclosure. Comparing such spaces could help distinguish broad occupational patterns from discourse patterns shaped by one platform's culture. It could also clarify which concerns are widely shared across the profession over time and across settings more broadly.

\section{Conclusion}

Professional online discourse among early childhood educators reflects a work environment shaped more by demands than by supports. In this corpus, educators most often discussed classroom pressures, workplace culture, regulation, compensation, and burnout, and demand-oriented posts carried more fear and negative affect than resource-oriented posts. These findings are consistent with the view that occupational strain in ECE is discussed as a structural rather than merely episodic feature of work. Large-scale text analysis cannot replace direct measures of well-being, but it can help illuminate how practitioners collectively describe and emotionally register the conditions of their work.

\section*{Competing Interests}
The author declares no competing interests.

\section*{Ethics Statement}
This study analyzed publicly available online material and did not involve direct interaction or intervention with human participants. No effort was made to identify users, reconstruct posting histories, or reproduce verbatim excerpts in ways that would increase traceability.

\section*{Data Availability}
The dataset was derived from publicly accessible Reddit posts. Because post text may contain potentially identifying language, raw data should be shared only in de-identified form and in ways consistent with platform terms and ethical norms for internet research.


\begin{thebibliography}{99}

\bibitem{agyapong2022} Agyapong, B., Obuobi-Donkor, G., Burback, L., \& Wei, Y. (2022). Stress, burnout, anxiety and depression among teachers: A scoping review. \textit{International Journal of Environmental Research and Public Health, 19}(17), 10706. \url{https://doi.org/10.3390/ijerph191710706}

\bibitem{ahmed2025} Ahmed, S. K., Mohammed, R. A., Nashwan, A. J., et al. (2025). Using thematic analysis in qualitative research. \textit{Journal of Medicine, Surgery, and Public Health, 6}, 100198. \url{https://doi.org/10.1016/j.glmedi.2025.100198}

\bibitem{bakker2007} Bakker, A. B., \& Demerouti, E. (2007). The job demands-resources model: State of the art. \textit{Journal of Managerial Psychology, 22}(3), 309--328. \url{https://doi.org/10.1108/02683940710733115}

\bibitem{carey2024} Carey, S., \& Sutton, A. (2024). Early childhood teachers' emotional labour: The role of job and personal resources in protecting well-being. \textit{Teaching and Teacher Education, 148}, 104699. \url{https://doi.org/10.1016/j.tate.2024.104699}

\bibitem{dan2025} Dan, E., Zhu, J., \& Jin, R. (2025). Exploring suicide factors in online discourse: Sentiment and thematic analysis of Reddit. \textit{ACM Transactions on the Web, 19}(4), Article 39. \url{https://doi.org/10.1145/3716546}

\bibitem{dechoudhury2013} De Choudhury, M., Counts, S., \& Horvitz, E. (2013). Social media as a measurement tool of depression in populations. In \textit{Proceedings of the 5th Annual ACM Web Science Conference} (pp. 47--56). Association for Computing Machinery. \url{https://doi.org/10.1145/2464464.2464480}

\bibitem{demerouti2001} Demerouti, E., Bakker, A. B., Nachreiner, F., \& Schaufeli, W. B. (2001). The job demands-resources model of burnout. \textit{Journal of Applied Psychology, 86}(3), 499--512. \url{https://doi.org/10.1037/0021-9010.86.3.499}

\bibitem{dickerson2025} Dickerson, M. K., Fenech, M., \& Stratigos, T. (2025). Working with families: An investigation of early childhood educators' emotional labour and wellbeing. \textit{Early Years, 45}(3--4), 593--609. \url{https://doi.org/10.1080/09575146.2024.2393143}

\bibitem{fuentes2025} Fuentes-Vilugron, G., Baeza-Vargas, Y., Fuentes-Fuentes, D., et al. (2025). Mental health and burnout levels of early childhood education pedagogical teams in Chile. \textit{Frontiers in Education, 10}, 1680412. \url{https://doi.org/10.3389/feduc.2025.1680412}

\bibitem{grant2019} Grant, A. A., Jeon, L., \& Buettner, C. K. (2019). Relating early childhood teachers' working conditions and well-being to their turnover intentions. \textit{Educational Psychology, 39}(3), 294--312. \url{https://doi.org/10.1080/01443410.2018.1543856}

\bibitem{hartmann2022} Hartmann, J. (2022, October 2). \textit{j-hartmann/emotion-english-distilroberta-base}. Hugging Face. \url{https://huggingface.co/j-hartmann/emotion-english-distilroberta-base}

\bibitem{jogi2023} Jogi, A., Aulen, A., Pakarinen, E., \& Lerkkanen, M. (2023). Teachers' daily physiological stress and positive affect in relation to their general occupational well-being. \textit{British Journal of Educational Psychology, 93}(1), 368--385. \url{https://doi.org/10.1111/bjep.12561}

\bibitem{madigan2021} Madigan, D. J., \& Kim, L. E. (2021). Towards an understanding of teacher attrition: A meta-analysis of burnout, job satisfaction, and teachers' intentions to quit. \textit{Teaching and Teacher Education, 105}, 103425. \url{https://doi.org/10.1016/j.tate.2021.103425}

\bibitem{oosterhoff2020} Oosterhoff, A., Oenema-Mostert, I., \& Minnaert, A. (2020). Constrained or sustained by demands? Perceptions of professional autonomy in early childhood education. \textit{Contemporary Issues in Early Childhood, 21}(2), 138--152. \url{https://doi.org/10.1177/1463949120929464}

\bibitem{purper2023} Purper, C. J., Thai, Y., Frederick, T. V., \& Farris, S. (2023). Exploring the challenge of teachers' emotional labor in early childhood settings. \textit{Early Childhood Education Journal, 51}(4), 781--789. \url{https://doi.org/10.1007/s10643-022-01345-y}

\bibitem{schaufeli2014} Schaufeli, W. B., \& Taris, T. W. (2014). A critical review of the Job Demands-Resources Model: Implications for improving work and health. In G. F. Bauer \& O. Hammig (Eds.), \textit{Bridging occupational, organizational and public health} (pp. 43--68). Springer. \url{https://doi.org/10.1007/978-94-007-5640-3_4}

\bibitem{schlueter2024} Schlueter, L., McGee, D., Link, T., Badanes, L., Dmitrieva, J., \& Watamura, S. (2024). Physiologic stress in the classroom: Does teacher's cortisol expression influence children's afternoon rise in cortisol at childcare? \textit{Psychology in the Schools, 61}, 2240--2254. \url{https://doi.org/10.1002/pits.23163}

\bibitem{stein2022} Stein, R., Garay, M., \& Nguyen, A. (2022). It matters: Early childhood mental health, educator stress, and burnout. \textit{Early Childhood Education Journal}. Advance online publication. \url{https://doi.org/10.1007/s10643-022-01438-8}

\bibitem{tummers2021} Tummers, L. G., \& Bakker, A. B. (2021). Leadership and Job Demands-Resources Theory: A systematic review. \textit{Frontiers in Psychology, 12}, 722080. \url{https://doi.org/10.3389/fpsyg.2021.722080}

\bibitem{wiltshire2024} Wiltshire, C. A. (2024). Early childhood education teacher workforce: Stress in relation to identity and choices. \textit{Early Childhood Education Journal, 52}(4), 655--668. \url{https://doi.org/10.1007/s10643-023-01468-w}

\bibitem{zhu2023} Zhu, J., Yalamanchi, N., Jin, R., Kenne, D. R., \& Phan, N. (2023). Investigating COVID-19's impact on mental health: Trend and thematic analysis of Reddit users' discourse. \textit{Journal of Medical Internet Research, 25}, e46867. \url{https://doi.org/10.2196/46867}

\bibitem{zhu2025probing} Zhu, J., Zhang, X., Jin, R., Jiang, H., \& Kenne, D. R. (2025). Probing public perceptions of antidepressants on social media: Mixed methods study. \textit{JMIR Formative Research, 9}(1), e62680.

\bibitem{zhu2025emotions} Zhu, J., Jiang, H., Wang, Y., Coifman, K. G., Jin, R., \& Kenne, D. R. (2025). Emotions, context, and substance use in adolescents: A large language model analysis of Reddit posts. \textit{arXiv preprint arXiv:2501.14037}.

\bibitem{zhu2025leveraging} Zhu, J., Jin, R., Jiang, H., Wang, Y., Zhang, X., \& Coifman, K. G. (2025). Leveraging large language models to analyze emotional and contextual drivers of teen substance use in online discussions. \textit{arXiv e-prints}, arXiv--2501.

\end{thebibliography}
\end{document}